# Multiferroic properties of $Bi_{0.9-x}La_{0.1}Er_xFeO_3$ ceramics

**Abstract:**

Structural, electrical and magnetic properties of $Bi_{0.9-x}La_{0.1}Er_xFeO_3$ ($BLEFO_x$) (x = 0.05, 0.07, 0.1) polycrystalline ceramics prepared by solid solution route were studied. A phase transition from rhombohedral phase to monoclinic phase was observed for x=0.05-0.1 in ($BLEFO_x$). We have measured phase transition temperatures both α-β transition and low-Temperature (low-T) transitions in doped $BiFeO_3$. The transition peak near to $835^0$C corresponds to α to β-phase transition of $BiFeO_3$ were measured using diffential thermal analysis (DTA). Dielectric measurements shows the low-T transitions in $BLEFO_x$ (x = 0.05-0.1). Relatively high remanent magnetization of 0.1178emu/gm at 8T was observed in $BLEFO_x$ (x = 0.1).

**Pragya Pandit, S. Satapathy [a)], Poorva Sharma and P. K. Gupta**

Laser Materials Development & Device Division, Raja Ramanna Centre for Advanced Technology, Indore 452013, India

**S. M. Yusuf**

Solid State Physics Division, Bhabha Atomic Research Centre, Mumbai 400 085, India

[a)] **Address for correspondence:**

**E- mail:** srinu73@cat.ernet.in

Ph: 91 731 2488660

Fax: 91 731 2488650

Magnetoelectric multiferroics are attracting attention for both fundamental physics due to their unique coupling behaviour between ferroelectricity, ferromagnetism and ferroelasticity[1] and also because of their promising applications for devices in spintronics, information storage, sensing and actuation[2]. Due to the coexistence of ferroelectricity and magnetism in multiferroic materials, an external magnetic field can induce electric polarization and an external electric field can induce changes in magnetization. The most common means of achieving ferromagnetism is rarely compatible with ferroelectricity because the requirement of localized transition metal d electrons for magnetism is not compatible with second order Jahn Teller effects which requires empty d orbitals for ferroelectricity[3].

Further most of the known multiferroics are ferroelectric-antiferromagnetic exhibiting ME (Magnetoelectric effect) coupling well below room temperature[4]. $BiFeO_3$ ceramic with a rhombohedrally distorted perovskite structure[5-7] is an interesting candidate due to its higher ferroelectric curie temperature (1083 K)[8] and antiferromagnetism below Neel temperature ($T_N$ = 643 K)[5,6]. $BiFeO_3$ is a commensurate ferroelectric[8] and an incommensurate antiferromagnet[9] at room temperature. In this distorted structure, the hexagonal $[001]_H$ direction is equivalent to the pseudo-cubic $[111]_C$ direction, along which there is a three fold rotation and about which the $Bi^{+3}$ are displaced from their centrosymmetric positions. This distortion is polar and results in a $P_s$ orientation along $(111)_C$. The spontaneous polarization of 6.1 μC/cm$^2$ along [111] direction and 3.5μC/cm$^2$ along [100] direction are reported in $BiFeO_3$ single crystals[8]. Recent studies shows the single crystalline thin layers fabricated by pulsed laser deposition have shown a much higher spontaneous polarization along $[001]_c$ and $[111]_c$, approaching values of $0.6 C/m^2$ and $1 C/m^2$ at room temperature respectively[10, 11]. Large polarization of $12.75 \mu C/cm^2$ at 155kV/cm is observed in quenched $BiFeO_3$ ceramic[12].

In $BiFeO_3$ the spin is provided by transition metal cation $Fe^{+3}$. Spins in the neighboring atoms are antiparallel, leading to an antiferromagnetic ordering of the G-type. In this arrangement the $Fe^{+3}$ cations are surrounded by six nearest $Fe^{+3}$ neighbors, with opposite spin directions. Microscopically, the antiferromagnetic spin order is inhomogeneous for $BiFeO_3$. Precise neutron –diffraction studies have revealed that in bulk $BiFeO_3$, the G-type antiferromagnetic ordering is modulated by a long wave length $\lambda$= 620Å spiral spin structure which would lead to a cancellation of the magnetization[9] on a macroscopic scale. The cycloidal spin structure has been directed along $[110]_H$. Room temperature ferromagnetism may exist in $BiFeO_3$ if there is a continuing collapse of the space modulated spin structure. By applying higher magnetic field (>18T) the cycloid spin structure converted to homogenous spin order[13]. Destruction of spin cycloid is also reported in strained epitaxial oriented $(111)_c$ $BiFeO_3$ thin film[11]. Recently M.K.Singh[14] have tried to resolve the spin wave structure of $BiFeO_3$ based on the model of magnon in the spiral structure of $BiFeO_3$[15] and suggested a magnon reorientation transition at 140K similar to orthoferrites. Another problem to be counteracted in bulk $BiFeO_3$ is basically low resistivity[16]. Rare earth substitutions cause local structural distortions thus modifying their resistive and ferroelectric behaviour. A-site (Bi site) and B-site (Fe site) rare earth substitutions are known to result both in enhanced resistivity[17-18] and suppression of inhomogeneous cycloid spin structure[10, 19]. Enhancement of magnetization has been reported in $Bi_{1-x}R_xFeO_3$ and $BiFe_xR_{1-x}O_3$ ($R=Nd^{+3}, Tb^{+3}, Mn^{+2}$)[20, 21]. In $Bi_{1-x}Nd_xFeO_3$ there is a continuing collapse of space modulated spin structure, hence room temperature remanent magnetization (0.227emu/g) is observed in $Bi_{0.8}Nd_{0.2}FeO_3$[22]. Similarly the multiferroic properties have been enhanced for La doping in $BiFeO_3$[23]. V. R. Palkaret al.[24] has also reported the doping of Tb & La and discovered the coexistence of ferroelectricity and magnetism along with high dielectric constant and magnetoelectric coupling at room temperature.

In this communication we report the detailed study of Er and La doped $BiFeO_3$ ceramics. Four samples with the composition $BiFeO_3$ and $Bi_{0.9-x}La_{0.1}Er_xFeO_3$ ($BLEFO_x$) (x = 0.05, 0.07, 0.1) were prepared through solid state reaction. The structural transitions of doped $BiFeO_3$ samples have been described. Both metal-insulator transition and low - T transitions in doped $BiFeO_3$ have been measured. The enhancement of magnetization was observed in Er doped samples compared to pure $BiFeO_3$.

The conventional solid solution route method of wet mixing the oxides $Bi_2O_3$ (99.99%), $Fe_2O_3$ (99.99%), $La_2O_3$ (99.99%) and $Er_2O_3$ (99.99%) using isopropyl alcohol medium in stochiometric proportions was employed to prepare $Bi_{0.9-x}La_{0.1}Er_xFeO_3$ (x=0.05, 0.07, 0.1) samples. Small amount of La was added to stabilize the perovskite phase[9]. The powder was doubly calcined consecutively at $650^0$ C for 1 hour and $810^0$ C for 1.5 hours to achieve the desired phase. The calcined powder was leached in concentrated $HNO_3$ to wash off impurity phases. The leached powders were mechanically pressed under 100 MPa and sintered at $850^o$ C for 7 hours.

Crystalline phases of the sintered pellets were identified using X-ray Diffractometer (Rigaku). Calorimetric properties have been measured using Differential Thermal Analysis (SETARAM model no TG/DTA-92B) to determine Curie temperature and magnetic transition temperature. Dielectric measurements were carried out in the frequency range (100 Hz - 10 KHz) using impedance analyzer (HP 4194 A). Magnetic measurements (M-H) loop were performed using vibrating sample magnetometer (VSM).

Fig. 1 shows the x-ray diffraction patterns of pure $BiFeO_3$ and $Bi_{(0.9-x)}La_{0.1}Er_xFeO_3$ (x = 0.05, 0.07 and 0.1). A single-phase rhombohedrally distorted structure R3C is observed with no trace of other (or impurity) phase within the uncertainity of XRD in case of pure $BiFeO_3$ [24, 25] (fig. 1(a)). The XRD patterns are in excellent accord with the powder data of JCPDS Card No. 71-2494. The lattice parameters for pure $BiFeO_3$ hexagonal unit cell are calculated to be a=5.5778$A^0$ c=13.8685 $A^0$ and the volume of unit cell is 373.6700 $(A^o)^3$. The fig. 1(b)-(d) depict the XRD pattern of $BLEFO_x$ (x = 0.05, 0.07 and 0.1 respectively). The magnified patterns of pure $BiFeO_3$ samples in the vicinity of $2\theta = 32^o$ shows that the (104) and (110) diffraction peaks are completely separate in a pure $BiFeO_3$, but the peaks are shifted and overlapped to a single peak when Bi atoms are substituted by Er[26]. This result implies that the rhombohedral structure is distorted to a monoclinic or orthorhombic structure by Er doping. After indexing and refinement it is clear that after La and Er doping the crystal structure of $BiFeO_3$ changes from rhombohedral (R3C) to monoclinic (C2). From XRD, second phase formation is not evident for x = 0.05, 0.07 and 0.1 in $BLEFO_x$.

Fig. 2 shows the DTA curves of pure $BiFeO_3$ and $BLEFO_x$ (x = 0.05, 0.07 and 0.1) samples for both cooling and heating cycles at the rate of $10^oC$/min. Sharp peak signifying first order phase transition could be identified in the heating and cooling cycles. This transition peak is similar to that observed by Palai et al.[27]. The transition peak near to $835^0C$ corresponds to α to β-phase transiton of $BiFeO_3$ [27]. The endothermic and exothermic peaks were observed at 835$^o$ C during heating and 808$^o$ C during cooling respectively for pure $BiFeO_3$. No signature of the magnetic transition was evident in the DTA diagram because of the small energy change associated with magnetic transition. Fig. 2(b)-2(d) clearly shows the decrease in Tc on subsequent Er doping for $BLEFO_x$ (x = 0.05, 0.07 and 0.1). The ferroelectric transition decreased (during heating) from 835$^0$C ($BiFeO_3$) to 792$^0$C (in case of 10 mol% doping of Er). Ion modification lowered the ferroelectric phase transition temperature which may be attributed to defects generated due to doping.

Fig. 3 shows the temperature variation of dielectric constant and dielectric loss tangent of $BiFeO_3$ and $BLEFO_x$ (x = 0.05, 0.07 and 0.1) samples within frequency range 100 Hz to 10 kHz. Fig. 3(a) and 3(b) show a peak at 338$^0$C corresponding to the Neel transition temperature of pure $BiFeO_3$. The dielectric anomaly at 835$^o$ C in case of pure $BiFeO_3$ corresponds to the ferroelectric phase transition as confirmed by the DTA measurements. The room temperature dielectric constant at 100Hz was increased from 50 for pure BFO to 8900 for $BLEFO_x$(x =0.05) at room temperature. Similarly the dielectric loss tangent decreases from 0.066 pure BFO to 0.039 for $BLEFO_x$(x =0.05) at room temperature at 100Hz.

But further addition of Er causes a reduction of dielectric constant i.e. [for $BLEFO_x$(x = 0.07) and for $BLEFO_x$(x = 0.1) dielectric constant is 3140 and 2000 respectively]. The dielectric constant value for $BLEFO_x$(x = 0.07 and 0.1) may not exact value of dielectric constant because further addition of Er causes an increase in dielectric loss (tan $\delta$) also i.e. [for $BLEFO_x$(x = 0.07) and for $BLEFO_x$(x = 0.1) losses are 0.071 and 0.823 respectively]. The dielectric constant has decreased with the increase in frequency for all of the samples due to difference in the contribution from polarizations [28, 29].

A dielectric anomaly was visible near the Neel transition temperature 338$^0$C which signifies magnetoelectric coupling[30]. The magnetoelectric coupling temperature which is observed in pure $BiFeO_3$ at 338$^0$C diffused and shifted to 185$^0$ C for $Bi_{0.85}Er_{0.05}La_{0.1}FeO_3$ (fig. 3(c) and fig. 3(d)) and was further diffused for $Bi_{0.83}Er_{0.07}La_{0.1}FeO_3$ (fig. 3(e) and fig. 3(f)). But no signature of this low-T transition temperature was visible in case of $Bi_{0.8}Er_{0.1}La_{0.1}FeO_3$ up to 350$^0$C (fig. 3(g) and fig. 3(h)). So for $Bi_{0.8}Er_{0.1}La_{0.1}FeO_3$ composition no magnetoelectric coupling was observed in dielectric data.

For the sample with pure phase BiFeO3, (figure 4(a)) a weak polarization of 0.02μC/cm$^2$ was observed under an applied field of 3kV/cm. No saturated hysteresis loop could be observed at room temperature under the above applied field. This implies that the samples are highly conductive at room temperature and only partial reversal of the polarization takes place quite similar to that observed by Teague et.al[8]. The relatively high conductivity of BFO is known to be attributed to the variable oxidation states of Fe ions ($Fe^{+2}$ to $Fe^{+3}$) which requires oxygen vacancies for charge compensation[2]. Also during synthesis, the slow heating rate and long sintering time will enable the equilibrium[8] concentration of the oxygen vacancies at high temperature to be reached and will result in the high oxygen vacancy concentration in the synthesized product. So the presence of $Fe^{+2}$ ions and oxygen deficiency leads to high conductivity. Due to high conductivity proper ferroelectric hysteresis loops were not measured at high fields for BLEFOx (x = 0.05-0.1) (Fig. 4(b)- (d)).

Fig. 5 shows the magnetization hysteresis (M-H) loops of $BLEFO_{x = 0 - 0.1}$ samples for the maximum magnetic field ($H_m$) of 8Tesla. It is clear that intrinsic antiferromagnetism is observed in $BLEFO_{x=0}$ and weak ferromagnetism with magnetization ($M_r$) of 0.076emu/gm is observed in $BLEFO_{x= 0.05}$. The $M_r$ increases as Er doping percentage increases. In case of $BLEFO_{x=0.07}$ the $M_r$ increases to 0.1164emu/gm which further increases to 0.1178emu/gm in case of $BLEFO_{x= 0.1}$. But there is no smooth reversal of magnetization in case of $BLEFO_{x= 0.1}$, when magnetic field is reversed. The humps in M~H curve for $BLEFO_{x= 0.1}$ may arise due to pinning of domain walls. The increase in Mr results from continuous collapse of the space modulated spin structure of $BLEFO_x$.

Er doped $BiFeO_3$ polycrystalline ceramics have been prepared successfully by solid state reaction route. The crystal structure of $Bi_{(0.9-x)} Er_xLa_{0.1}FeO_3$ changes from rhombohedral to monoclinic for $BLEFO_x$ (x = 0.05, 0.07 and 0.1). Both metal-insulator transition and low-T transition of Er doped $BiFeO_3$ have been observed. A dielectric anomaly was visible in the vicinity of Neel's transition in pure $BiFeO_3$ indicating coupling between the electric and magnetic order. These ceramics demonstrated an enhanced remanent magnetization in the range from 0.076emu/gm for x = 0.05 to 0.1178emu/gm for x=0.07 and magnetic hysteresis loops were also observed. We have not demonstrated ferroelectric switching because the conductivity is too high.

We are indebted to Prof. J. F. Scott for valuable suggestions.

The authors acknowledge the help received from Dr. S. B. Roy, Dr. Gurvinderjit Singh, Mr. Rajeev Bhatt and Mr. Indranil Bhaumik for characterization.

**Caption of Figures**

**Fig. 1**. XRD patterns of (a) $BLEFO_{x=0}$, (b) $BLEFO_{x=0.05}$ (c) $BLEFO_{x=0.07}$ and (d) $BLEFO_{x=0.1}$ ceramic samples.

**Fig. 2.** DTA thermograms of (a) $BLEFO_{x=0}$, (b) $BLEFO_{x=0.05}$ (c) $BLEFO_{x=0.07}$ and (d) $BLEFO_{x=0.1}$.

**Fig. 3.** Variation of dielectric constant and loss with temperature of (a-b) $BLEFO_{x=0}$, (c-d) $BLEFO_{x=0.05}$, (e-f) $BLEFO_{x=0.07}$ and (g-h) $BLEFO_{x=0.1}$ at different frequencies (100Hz to 10 kHz).

**Fig. 4**. Polarization hysteresis (*P*-*E*) loops of a) $BLEFO_{x=0}$, (b) $BLEFO_{x=0.05}$, (c) $BLEFO_{x=0.07}$ and (d) $BLEFO_{x=0.1}$ samples at room temperature

**Fig. 5.** Magnetization hysteresis (M-H) loops of pure $BiFeO_3$ and $BLEFO_x$ (x = 0.05, 0.07 and 0.1) at room temperature.

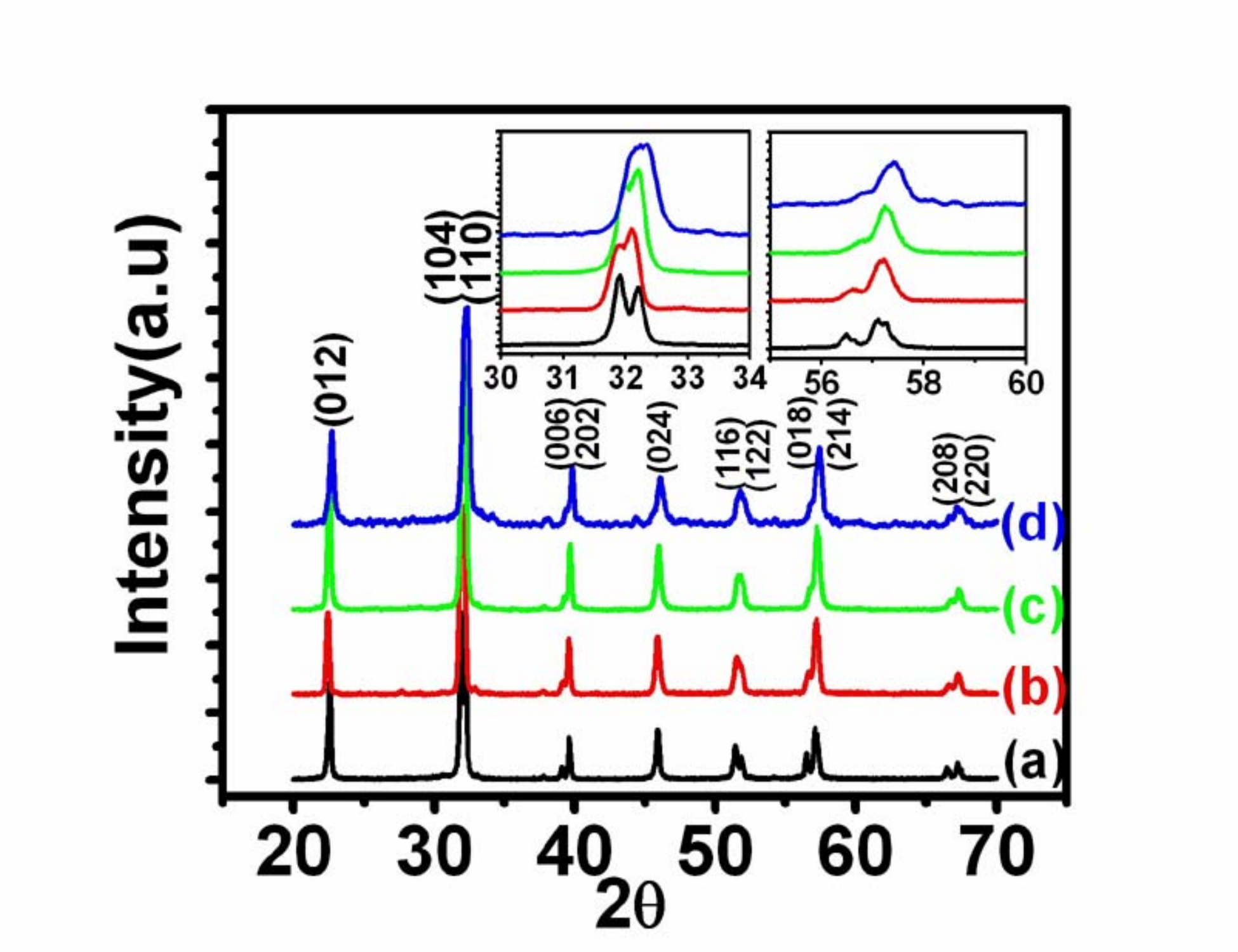

**Fig. 1**

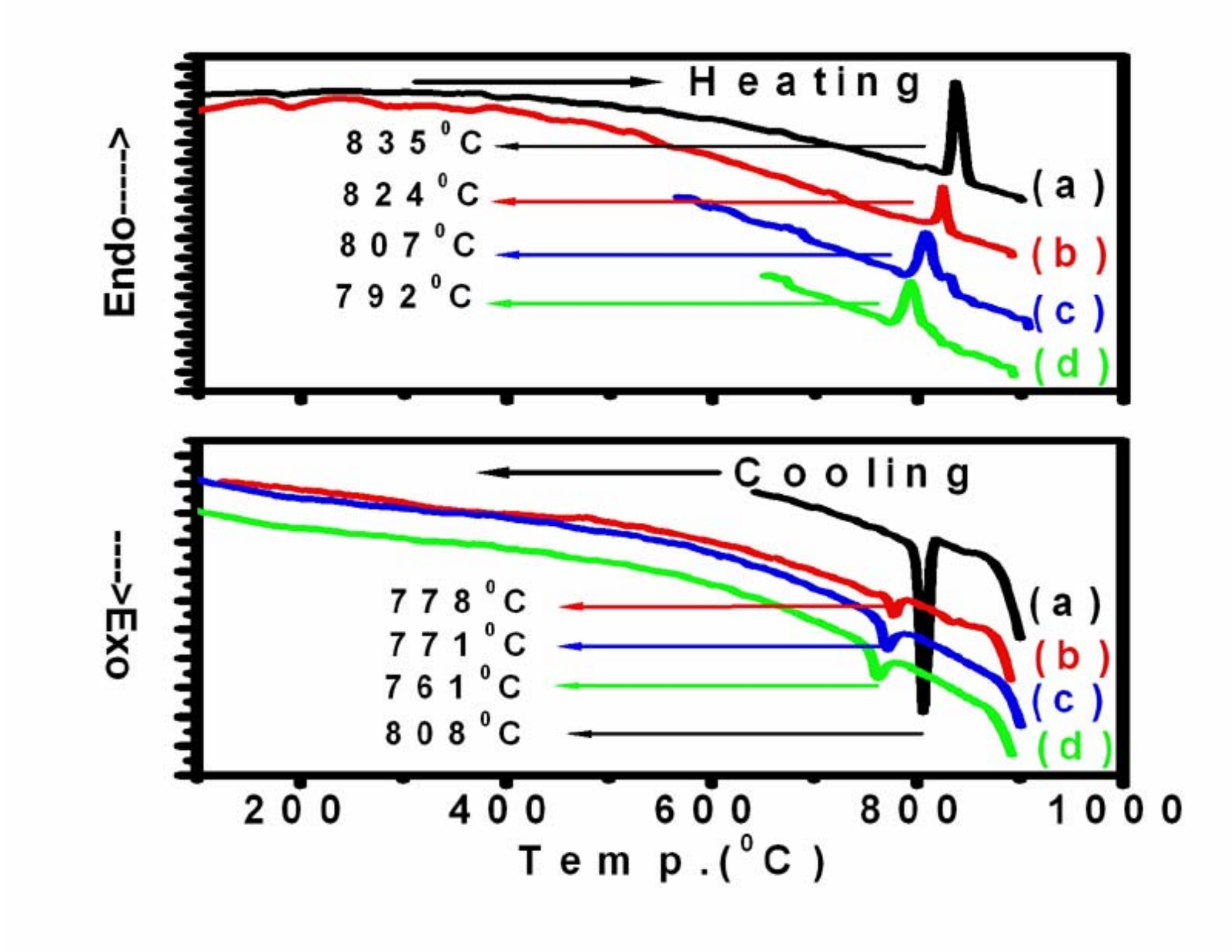

**Fig. 2**

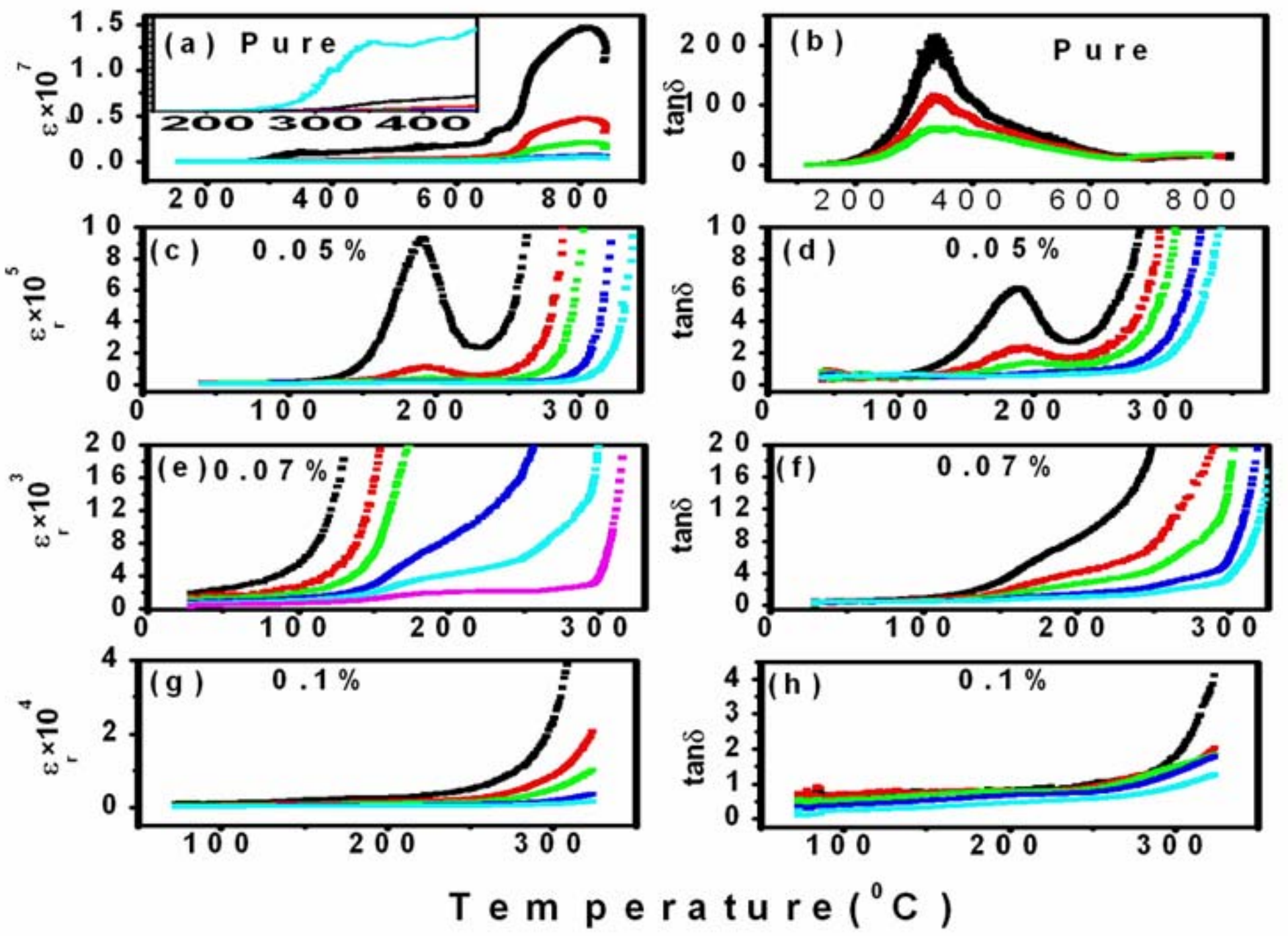

**Fig. 3**

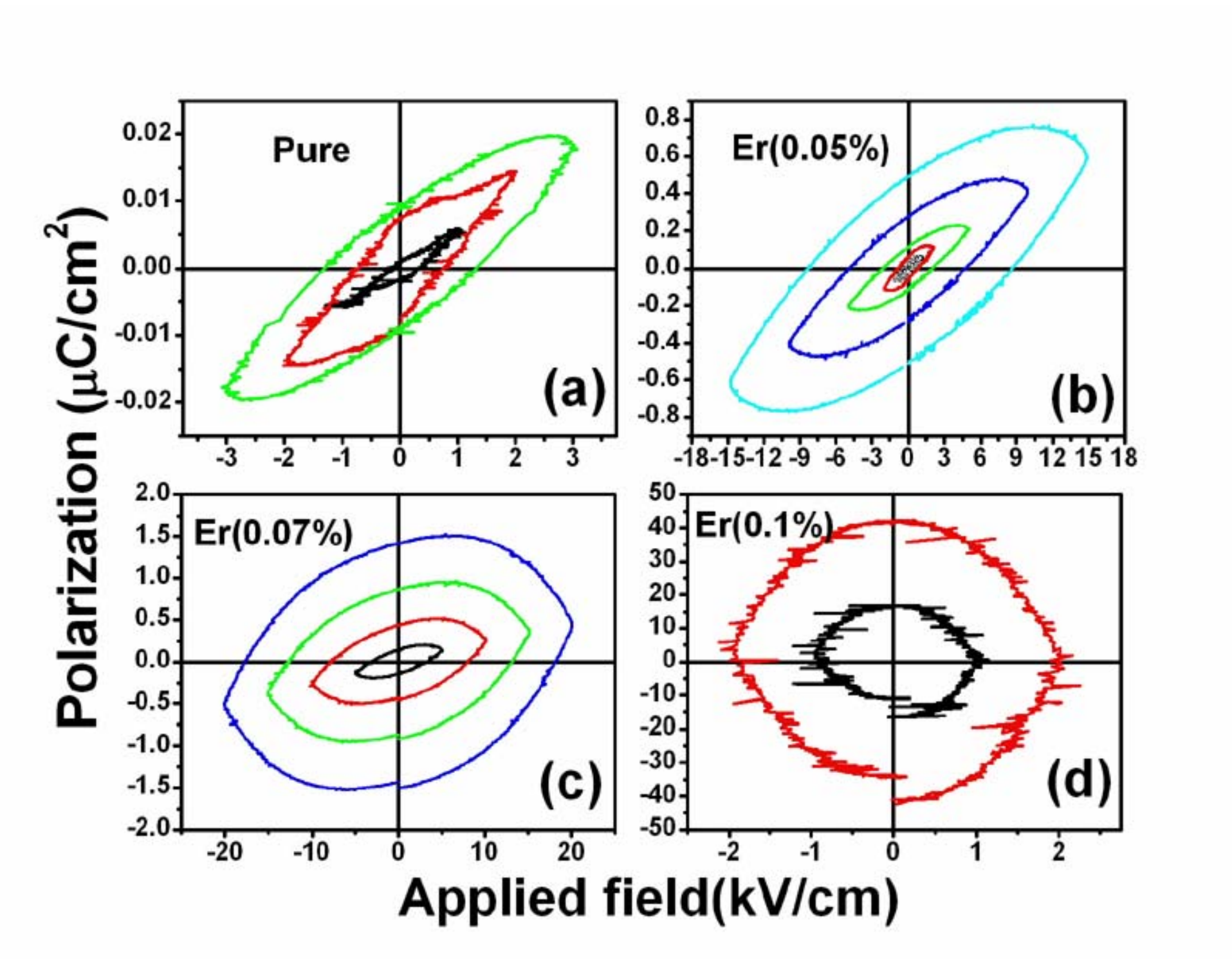

**Fig. 4**

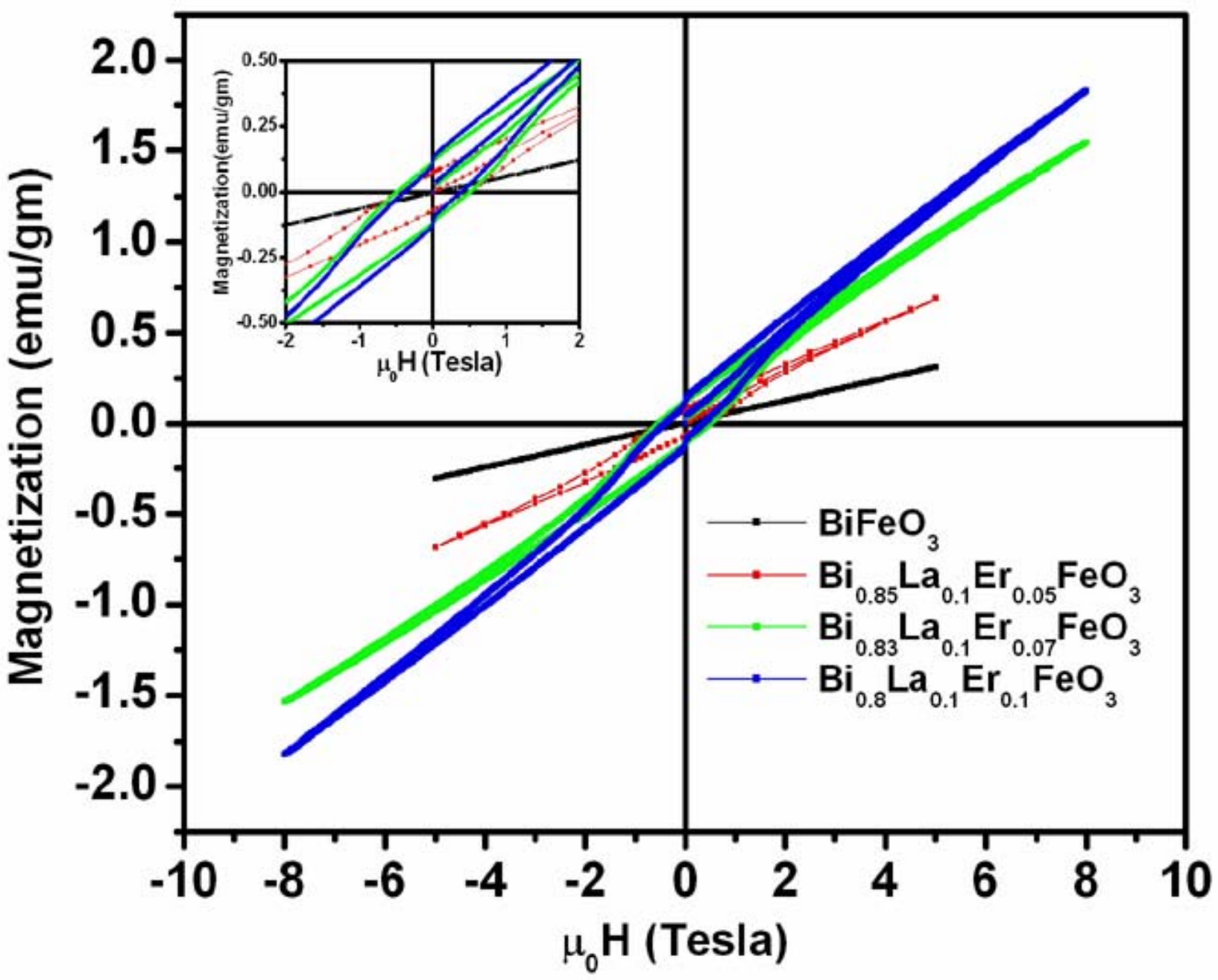

**Fig. 5**